\documentclass[twocolumn,showpacs,showkeys,amsmath,amssymb,superscriptaddress,floatfix,nofootinbib]{revtex4}

\usepackage{graphicx}
\usepackage{bm}
\usepackage{subfigure}

\def\vec#1{\mathchoice{\mbox{\boldmath$\displaystyle#1$}}
{\mbox{\boldmath$\textstyle#1$}}
{\mbox{\boldmath$\scriptstyle#1$}}
{\mbox{\boldmath$\scriptscriptstyle#1$}}}
\makeatletter

\begin{document}

\title{Two-body nucleon-nucleon correlations in Glauber models of relativistic heavy-ion collisions%
\footnote{Supported in part by the Polish Ministry of Science and Higher
Education, grants N202~034~32/0918 and N202~249235.}}

\author{Wojciech Broniowski}
\email{Wojciech.Broniowski@ifj.edu.pl}
\affiliation{The H. Niewodnicza\'nski Institute of Nuclear Physics, Polish Academy of Sciences, PL-31342 Krak\'ow, Poland}
\affiliation{Institute of Physics, Jan Kochanowski University, PL-25406~Kielce, Poland}

\author{Maciej Rybczy\'nski}
\email{Maciej.Rybczynski@ujk.edu.pl}
\affiliation{Institute of Physics, Jan Kochanowski University, PL-25406~Kielce, Poland}

\date{4 March 2010}

\begin{abstract}
We investigate the influence of the central two-body nucleon-nucleon correlations on several quantities observed in relativistic heavy-ion
collisions. It is demonstrated with explicit Monte Carlo simulations, that the basic correlation measures observed in relativistic
heavy-ion collisions, such as the fluctuations of participant eccentricity, initial size fluctuations, or the fluctuations of the  
number of sources producing particles, are all  
sensitive to the inclusion of the two-body correlations. The effect is at the level of about 10-20\%. Moreover, the realistic (Gaussian)
correlation function gives indistinguishable results from the hard-core repulsion, with the expulsion distance set to 0.9~fm. Thus,
we verify that for investigations of the considered correlation measures, it is sufficient to use the Monte Carlo
generators accounting for the hard-core repulsion.
\end{abstract}



\maketitle

\section{Introduction}

The atomic nucleus is closer to a self-bound saturated liquid than to a Fermi gas of non-interacting particles, as is 
for simplicity frequently assumed in studies of relativistic heavy-ion collisions. Thus the inclusion of correlations in the
initial configuration of nucleons in the colliding nuclei is {\em a priori} very important.
Recently Alvioli, Drescher, and Strikman~\cite{Alvioli:2009ab,Alvioli:page} 
generated distributions of nucleons in nuclei which account for the central
two-body nucleon-nucleon (NN) correlations. The procedure, based on the Metropolis search for configurations
satisfying constraints imposed by the NN correlations, reproduces the one-body Woods-Saxon distributions, as well as
central NN correlations, taken in the Gaussian form. This calculation is a
very important step in the investigations using the Glauber approach \cite{glauber,Czyz:1969jg} 
to relativistic heavy-ion collisions, as it is well known \cite{Baym:1995cz,Heiselberg:2000fk} that 
correlations induce event-by-event fluctuations of the measured quantities.

The Glauber Monte Carlo codes~\cite{Wang:1991hta,Werner:1988jr,Broniowski:2007nz,Alver:2008aq}
(for a discussion of physics issues see Ref.~\cite{Broniowski:2007nz} and the review \cite{Miller:2007ri})
which model the early phase of the collision, have not been incorporating, for practical reasons, realistic NN correlations.
Instead, the hard-core expulsion, easy to implement, is used. In that method, centers of
nucleons, whose positions are randomly generated
according to the Woods-Saxon one-body distribution, are not allowed to be placed closer to one-another
than the expulsion distance $d \sim 1$~fm, which simulates the hard-core NN repulsion.
It is not a priori clear that the results obtained with the
realistic (Gaussian) and the hard-core
correlations should be the same for various correlation measures used in the heavy-ion studies. 
Moreover, it is not obvious what precise value of $d$ should be taken to make the simulations 
most realistic. 

The purpose of this paper is to investigate, with the help of explicit Glauber Monte-Carlo simulations by {\tt GLISSANDO}
\cite{Broniowski:2007nz}, the role of the central two-body NN correlations for several
popular observables in relativistic
heavy-ion collisions. In particular, we look at the
following {\em fluctuation measures}: the
participant eccentricity fluctuations related to the fluctuations of the elliptic
flow~\cite{Aguiar:2000hw,Miller:2003kd,Bhalerao:2005mm,Manly:2005zy,Andrade:2006yh,Voloshin:2006gz,Alver:2006pn,%
Alver:2006wh,Sorensen:2006nw,Broniowski:2007ft,Alver:2007rm,Hama:2007dq,Voloshin:2007pc}, the multiplicity
fluctuations as analyzed in the set up of the CERN NA49 experiment~\cite{Alt:2006jr},
and the recently investigated initial size fluctuations \cite{Broniowski:2009fm}, which influence the transverse-momentum
fluctuations~\cite{Gazdzicki:1992ri,Stodolsky:1995ds,Shuryak:1997yj,Mrowczynski:1997kz,%
Voloshin:1999yf,Baym:1999up,Appelshauser:1999ft,Voloshin:2001ei,Prindle:2006zz,Mrowczynski:2009wk,%
Adams:2003uw,Adamova:2003pz,Adler:2003xq,Adams:2005ka,Grebieszkow:2007xz,na49:2008vb}.
We find that all these measures are sensitive to the inclusion
of the two-body correlations at a level of about 10-20\%. However, the realistic (Gaussian)
correlation function gives virtually indistinguishable results from the calculations with the
hard-core repulsion, with the expulsion distance tuned to $d=0.9$~fm. Thus, we will argue that for
all practical terms of modeling the Glauber initial phase of the collision, it is sufficient to use the Monte Carlo
generators with the hard-core repulsion.

Certainly, the method of Ref.~\cite{Alvioli:2009ab} is more general, as it allows to include correlations from attractive
forces, as well as introduce the isospin dependence. These were recently considered in Ref.~\cite{Alvioli:2009ev},
and when these distributions are published, they can be implemented in Glauber generators and tested in a similar way as
in the present work.

\section{Nuclear correlations}

\begin{figure}[tb]
\begin{center}
\includegraphics[angle=0,width=0.46 \textwidth]{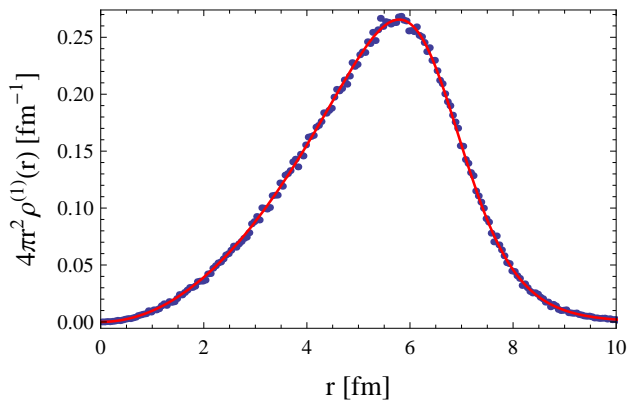}
\end{center}
\vspace{-6mm}
\caption{(color online). Radial one-body density of centers of nucleons
in $^{208}$Pb, $4\pi r^2 \rho^{(1)}(r)$, obtained numerically from the distributions of Ref.~\cite{Alvioli:2009ab,Alvioli:page}
(dots). The line shows the Woods-Saxon fit with the optimum parameters (\ref{param1}). \label{fig:oneb}}
\end{figure}

The method of Ref.~\cite{Alvioli:2009ab} imposes a given form of one- and two-body
nucleon distributions.
The one-body density is parametrized with the standard Woods-Saxon form
\begin{eqnarray}
\rho^{(1)}(r)=\frac{A}{1+e^{\frac{r-R}{a}}}.
\end{eqnarray}
Our fit to the distributions for $^{208}$Pb from~\cite{Alvioli:page} yields the
optimum parameters
\begin{eqnarray}
R= 6.59(1){\rm ~fm},\;\;\; a= 0.549(2){\rm ~fm}, \label{param1}
\end{eqnarray}
where the uncertainties follow from the regression analysis on
the available sample \cite{Alvioli:page} of $10^5$ configurations.
The result of our numerical simulation
is displayed in Fig.~\ref{fig:oneb}.

\begin{figure}[tb]
\begin{center}
\includegraphics[angle=0,width=0.46 \textwidth]{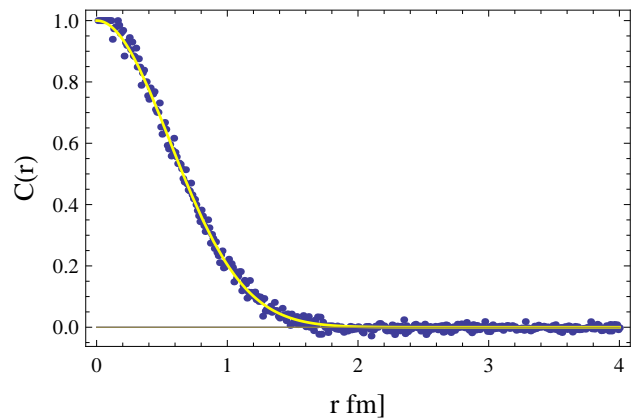}
\end{center}
\vspace{-6mm}
\caption{(color online). Central two-body NN radial correlation density for $^{208}$Pb, obtained from Eq.~(\ref{C:corr}) (points),
and the Gaussian fit of Eq.~(\ref{gauss},\ref{par:b}) (line). \label{fig:corr}}
\end{figure}

The radial two-body correlation function $C(r)$ is defined as \cite{Alvioli:2009ab}
\begin{eqnarray}
C(r)=1-\frac{\rho^{(2)}_C(r)}{\rho^{(2)}_U(r)}, \label{C:corr}
\end{eqnarray}
where $\rho^{(2)}_C(r)$ and $\rho^{(2)}_U(r)$ are the correlated and uncorrelated
radial two-body densities,
\begin{eqnarray}
\rho^{(2)}_i(r)=\int d^2 \Omega \int d^3 R \,\rho_i^{(2)}(\vec{R}+\vec{r}/2,\vec{R}-\vec{r}/2). \label{rho2}
\end{eqnarray}
Here $\rho^{(2)}_i(\vec{r_1},\vec{r_2})$, $i=C,U$, denotes the
appropriate two-nucleon density, $\vec{r}$ is the relative coordinate, $r=|\vec{r}|$,
and $\Omega$ corresponds to the two angles
associated with $\vec{r}$, over which the density is integrated. The correlated density is read off from the distributions
\cite{Alvioli:page} with the help of {\tt GLISSANDO} by histogramming
the relative distances between the centers of nucleons in the same nucleus, while the uncorrelated density is found by
taking the pairs of nucleons from {\em different} nuclei (this corresponds to the well-known {\em mixing technique}, which
gets rid of correlations). The result of our procedure is shown in Fig.~\ref{fig:corr}. We recover the Gaussian central
NN correlation, implemented in the procedure of Ref.~\cite{Alvioli:2009ab},
\begin{eqnarray}
C(r)=e^{-\frac{r^2}{2 b^2}}, \label{gauss}
\end{eqnarray}
with
\begin{eqnarray}
b= 0.561(1). \label{par:b}
\end{eqnarray}
The uncertainty comes from the finite sample of $10^5$ configurations from~\cite{Alvioli:page}.

Thus indeed the distributions of \cite{Alvioli:2009ab,Alvioli:page}
properly implement the one-body density and the Gaussian central two-body correlations.
The purpose of the above study was to read off the one-body parameters (\ref{param1}), which in the following sections
will be input in the
generation of the uncorrelated distributions by the Glauber simulations with {\tt GLISSANDO} \cite{Broniowski:2007nz}. Results from the
uncorrelated distributions will be compared to the correlated case, where the correlated distributions of
Ref.~\cite{Alvioli:2009ab} will be fed directly into our simulations.

\section{Glauber models}

The prototype Glauber model used in the heavy-ion phenomenology is the {\em wounded-nucleon model} \cite{Bialas:1976ed}.
A wounded nucleon has collided inelastically
at least once in the collision process. Variants of the approach
\cite{Kharzeev:2000ph,Back:2001xy,Back:2004dy,Broniowski:2007nz} admix a certain fraction of binary collisions
to the wounded nucleons, which
leads to a better overall description of multiplicities of the produced particles. In this {\em mixed model}, investigated 
in this work,
the number of the produced particles is proportional to the number of {\em sources}
\begin{eqnarray}
N_s=(1-\alpha)N_{w}/2+\alpha N_{\rm bin}, \label{ns}
\end{eqnarray}
where $N_w$ is the number
of the wounded nucleons, and $N_{\rm bin}$ the number of binary NN collisions. The fits to particle
multiplicities of Ref.~\cite{Back:2004dy} give $\alpha = 0.145$ for collisions at $\sqrt{s_{NN}}=200$~GeV, and $\alpha = 0.12$ for $\sqrt{s_{NN}}=19.6$~GeV. Extrapolation to the LHC energies yields $\alpha \simeq 0.2$. 

More sophisticated approaches
\cite{Bozek:2005eu,Werner:2007bf,Manninen:2008mg}
discriminate between the nucleons which have collided only once (corona) and more than once (core), which leads to an appealing
physical picture. Also, the wounded-quark model 
\cite{Bialas:1977en,Bialas:1980zw,Eremin:2003qn,Bialas:2006kw,Bialas:2007eg,Bzdak:2008gw} yields a successful phenomenology, 
All in all, the Glauber picture of the initial stage of the relativistic heavy-ion collision is a key element of many
phenomenological analyses of the particle production mechanism.

In this paper we apply the mixed model for the $^{208}$Pb-$^{208}$Pb collisions, with $\alpha=0.12$, corresponding to the
highest SPS energy. We term the locations of centers of the wounded nucleons or the binary collisions as ``sources'', with the weight of the
wounded nucleon $w_i=(1-\alpha)/2$, and the weight of the binary collision $w_i=\alpha$. A source
emits particles, according to a superposed distribution \cite{Broniowski:2007nz}.

While for the one-body measures, such as the particle multiplicities or spectra, only the one-body distributions matter
and correlations are irrelevant, the fluctuations measures are expected to be sensitive to the NN correlations in the
nucleon distributions. These are examined in detail in the next section.

\section{Results of simulations}

In this section we compare the results of the Glauber calculation initialized with the distributions of Ref.~\cite{Alvioli:2009ab,Alvioli:page}
(solid lines in the figures), with uncorrelated distributions (dashed lines), and with the distributions accounting for the hard-core repulsion
with the expulsion radius $d=0.9$~fm (dotted lines).
The simulations are performed with {\tt GLISSANDO} \cite{Broniowski:2007nz}. 

We note that in the
case with no correlations we simply use the Woods-Saxon parameters (\ref{param1}), while in the case with the hard-core repulsion we need to
start with a somewhat more compact distribution, as the expulsion leads to swelling, as explained in Ref.~\cite{Broniowski:2007nz}. We find that
starting the Monte Carlo generation with $R=6.44$~fm and $a=0.549$~fm, leads, with $d=0.9$~fm, to the one-body
distribution with parameter values (\ref{param1}). This construction, with shrunk ``bare'' one-body
distributions, is important, as that way all calculations presented in the
figures correspond to {\em identical} one-body distribution, and the differences in results are caused entirely by 
the two-body correlations.

\subsection{Eccentricity}

We start with a measure sensitive to the fluctuations, the so-called {\em participant} eccentricity.
This measure appears in the studies of the event-by-event fluctuations of the initial shape, in particular of its
elliptic component 
\cite{Aguiar:2000hw,Miller:2003kd,Bhalerao:2005mm,Manly:2005zy,Andrade:2006yh,Voloshin:2006gz,Alver:2006pn,%
Alver:2006wh,Sorensen:2006nw,Broniowski:2007ft,Alver:2007rm,Hama:2007dq,Voloshin:2007pc}. 
The effect is important, as the fluctuations lead to enhanced
eccentricity of the initial system, and as a result of the subsequent hydrodynamic evolution,
to enhanced elliptic flow. The participant eccentricity is defined in each event as
\begin{eqnarray}
\varepsilon^\ast=\frac{\sqrt{\left ( \sigma_x^2-\sigma_y^2 \right )^2+4\sigma_{xy}^2}}{\sigma_x^2+\sigma_y^2}, \label{epss}
\end{eqnarray}
where $\sigma^2_x$ and $\sigma^2_y$ are the variances of the two transverse coordinates, and $\sigma_{xy}$ is the
covariance. Specifically, in each event
\begin{eqnarray}
\langle x \rangle =\sum_i w_i x_i, \;\; \sigma^2_x = \sum_i w_i (x_i-\langle x \rangle)^2,
\end{eqnarray}
and similarly for the $y$ variable and the covariance. The index $i$ runs over all generated 
sources, and $w_i$ are the weights.
The quantity $\varepsilon^\ast$ has the interpretation of the eccentricity evaluated event-by-event in a {\em variable
reference frame} \cite{Broniowski:2007ft}, rotated in such a way that the eccentricity in a given event is maximized.

\begin{figure}[tb]
\begin{center}
\includegraphics[angle=0,width=0.485 \textwidth]{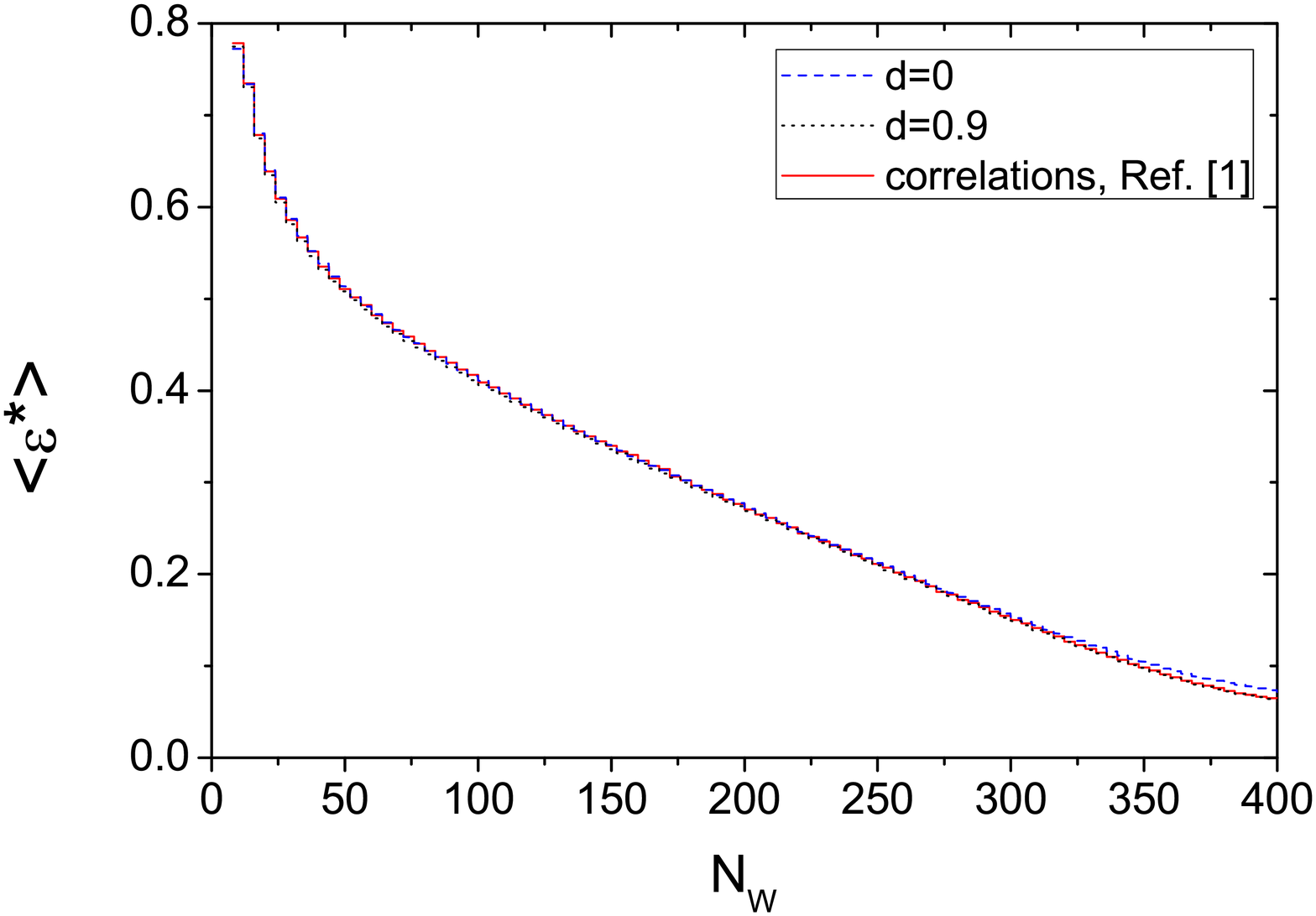}
\includegraphics[angle=0,width=0.485 \textwidth]{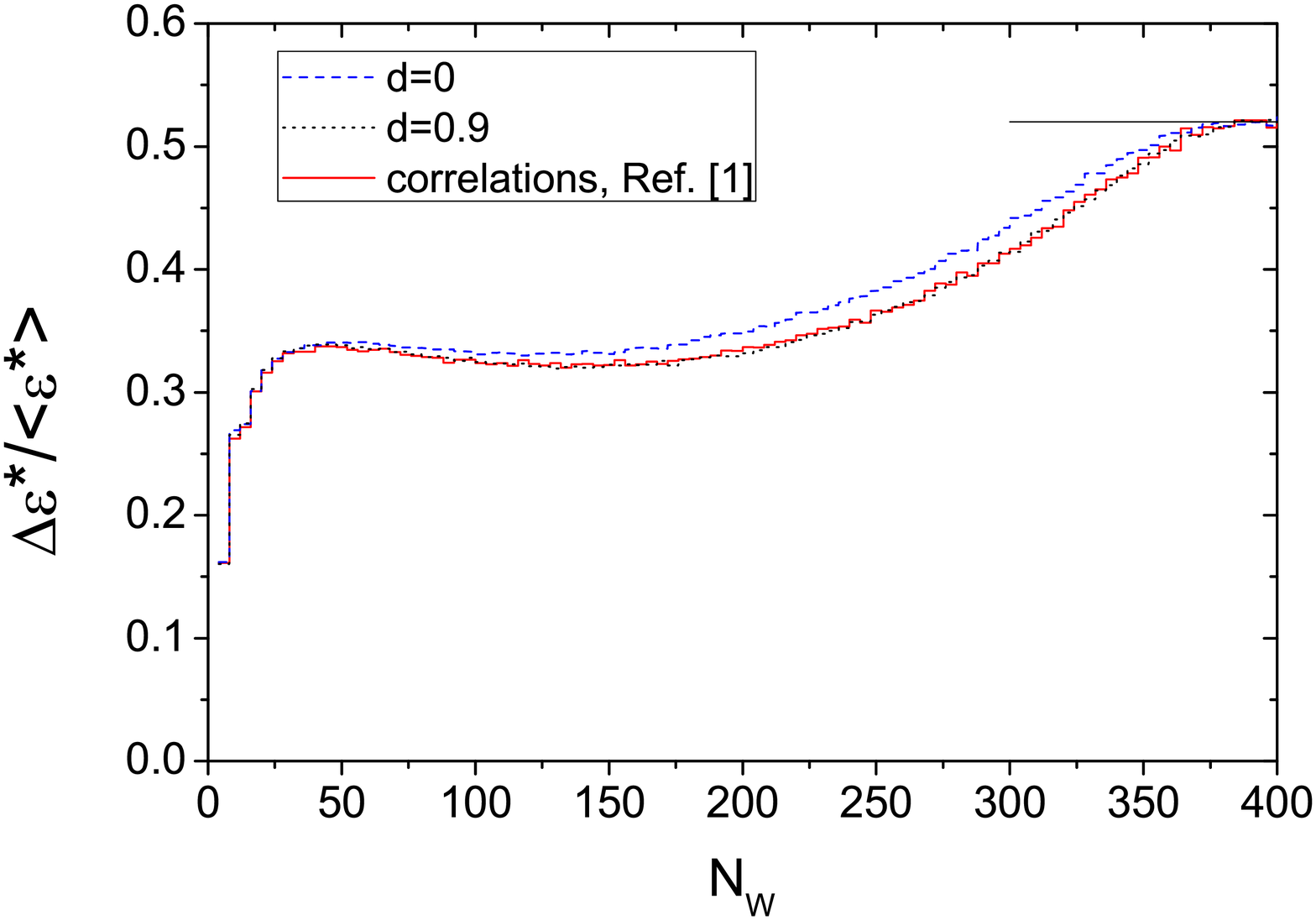}
\end{center}
\vspace{-6mm}
\caption{(color online). Top: the average participant eccentricity, $\langle \varepsilon^\ast \rangle$, vs the number of
wounded nucleons, $N_w$, obtained with the three investigated nucleon distributions described in the text.
Bottom: The scaled standard deviation $\Delta \varepsilon^\ast/\langle \varepsilon^\ast \rangle$ obtained from an event-by-event
study. The short horizontal line at the most central events corresponds to the theoretical value $\sqrt{4/\pi-1}$
of Ref.~\cite{Broniowski:2007ft} following from the central limit theorem. \label{fig:eps}}
\end{figure}

In the top panel of Fig.~\ref{fig:eps} we show the dependence of the event-by-event average,
$\langle \varepsilon^\ast \rangle$, on the number of wounded
nucleons (determining the {\em centrality} of the event). We note that the three calculations are virtually
indistinguishable, except for a tiny difference for the most central collisions, where the uncorrelated case is
a few percent higher. The same conclusions were reached in the analogous study of
eccentricity in Ref.~\cite{Tavares:2007mu}.

The bottom panel of Fig.~\ref{fig:eps} shows the scaled standard deviation,
$\Delta \varepsilon^\ast/\langle \varepsilon^\ast \rangle$, obtained from our event-by-event
analysis. We note a significant difference between the
uncorrelated case, which has up to 10\% larger fluctuations at intermediate
centralities, and the cases with correlations. 
However, the calculations with the realistic NN correlations and the hard-core
correlations give an indistinguishable result, with the two curves overlapping within the
statistical noise.

The short horizontal line at the most central events corresponds to the theoretical value $\sqrt{4/\pi-1}$
of Ref.~\cite{Broniowski:2007ft}, following from the central limit theorem.

\subsection{Multiplicity fluctuations}

Next, we consider a quantity relevant for the multiplicity fluctuations as measured in the NA49 experimental setup~\cite{Alt:2006jr}, where the
number of participants in the {\em projectile} is determined via the VETO calorimeter.
Significant fluctuations of the number of sources may follow in this case from the fact that even at a 
fixed number of the wounded nucleons in the projectile, the number of
wounded nucleons in the target fluctuates due to the statistical nature of the Glauber approach.
The fluctuations of multiplicity in nucleus-nucleus collisions were also investigated experimentally
in \cite{Albrecht:1989kh,Bachler:1992jp,Aggarwal:2001aa,Adare:2008ns}. 
We recall \cite{Rybczynski:2004zi,Rybczynski:2008zg} that the simple
superposition models with the effect of fluctuations of the target wounded nucleons
are not able to explain the data of Ref.~\cite{Alt:2006jr}. Nevertheless, for the present purpose
of analyzing the importance of the NN correlations, the effect serves its purpose.

\begin{figure}[tb]
\begin{center}
\includegraphics[angle=0,width=0.485 \textwidth]{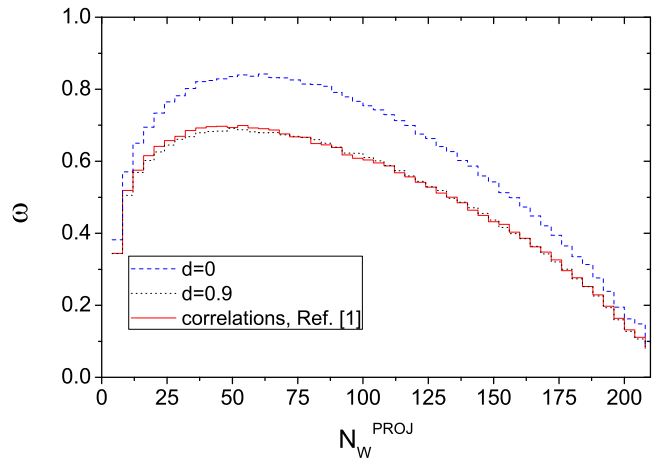}
\end{center}
\vspace{-6mm}
\caption{(color online). The scaled
variance of the number of sources, $\omega$,
plotted as a function of the wounded nucleons in the projectile, $N_w^{\rm PROJ}$. \label{fig:var}}
\end{figure}

In Fig.~\ref{fig:var} we show the scaled
variance of the {\em total} number of sources defined in Eq.~(\ref{ns}),
\begin{eqnarray}
\omega=\frac{{\rm var}(N_s)}{\langle N_s\rangle},
\end{eqnarray}
plotted as a function of the wounded nucleons in the projectile, $N_w^{\rm PROJ}$.
We note a significant, about 20\%, reduction of $\omega$ when the two-body NN correlations are included.
However, again there is no noticeable difference between the realistic (Gaussian) correlations and the hard-core
expulsion, as the two lower curves in the figure overlap.

\subsection{Size fluctuations}

\begin{figure}[tb]
\begin{center}
\includegraphics[angle=0,width=0.485 \textwidth]{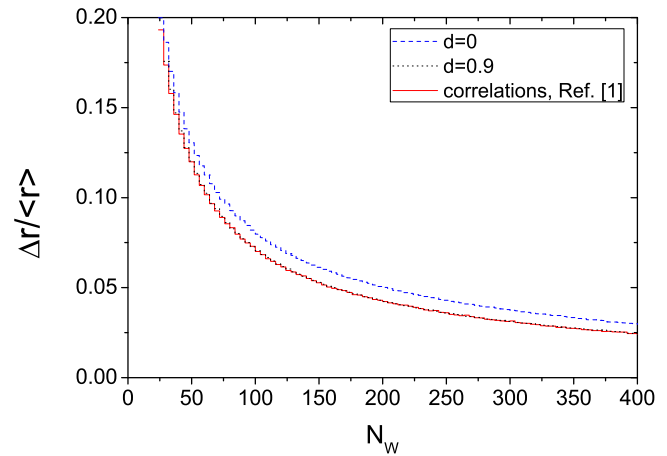}
\end{center}
\vspace{-6mm}
\caption{(color online). The scaled standard deviation of the size variable $r$ of Eq.~(\ref{rav}), plotted as a function
of the total number of wounded nucleons, $N_w$. \label{fig:size}}
\end{figure}

Finally, we look at the event-by-event {\em size} fluctuations, namely the fluctuations of the variable
\begin{eqnarray}
r =\sum_i w_i \sqrt{(x_i-\langle x \rangle)^2+(y_i-\langle y \rangle)^2}. \label{rav}
\end{eqnarray}
It was recently shown in Ref.~\cite{Broniowski:2009fm} that the initial size fluctuations are carried over via
hydrodynamics and statistical hadronization into the
event-by-event transverse-momentum fluctuations
\cite{Gazdzicki:1992ri,Stodolsky:1995ds,Shuryak:1997yj,Mrowczynski:1997kz,%
Voloshin:1999yf,Baym:1999up,Appelshauser:1999ft,Voloshin:2001ei,Prindle:2006zz,Mrowczynski:2009wk,%
Adams:2003uw,Adamova:2003pz,Adler:2003xq,Adams:2005ka,Grebieszkow:2007xz,na49:2008vb}, where they
lead to a natural description of the RHIC data for the measure $\sigma_{\rm dyn}(p_T)$.
In Fig.~\ref{fig:size} we show the scaled standard deviation of $r$.
Once again, the presence of the NN correlations reduces somewhat the fluctuations, while the realistic and
hard-core correlations with $d=0.9$~fm give virtually the same result.

\section{Conclusions}

We have checked by carrying out explicit Glauber Monte Carlo simulations with 
{\tt GLISSANDO} \cite{Broniowski:2007nz}, that the inclusion of the central NN correlations
influences the fluctuation measures in relativistic heavy-ion collisions at a level of, say, 10-20\%.
Comparison of the realistic (Gaussian) correlations implemented in Ref.~\cite{Alvioli:2009ab} and the hard-core correlations,
typically used in the Glauber Monte Carlo codes, shows that
they lead to the same results when the hard-core expulsion distance between the centers of nucleons is tuned to
\begin{eqnarray}
d=0.9~{\rm fm}. \label{eq:d}
\end{eqnarray}
Thus the main massage for the practitioners of the Glauber Monte Carlo models is that, at least for the investigated
observables, the implementation of the hard-core repulsion with $d$ given by Eq.~(\ref{eq:d}), straightforward to implement in Monte Carlo generators, leads to realistic predictions. We note that 
the dependence of the results on the value of $d$ is sensitive, as the excluded volume scales as $d^3$. 

Certainly, the general method of Ref.~\cite{Alvioli:2009ab} allows one to implement channel-dependent NN correlations, as well as
the nuclear attraction, relevant at intermediate distances. The role of these effects for the fluctuation measures in 
relativistic heavy-ion collisions can be investigated in a similar manner as in this work.
In essence, every effect which increases the ``regularity'' of the initial 
nucleon distributions of the colliding nuclei, 
such as the considered central NN correlations,
will have the tendency of decreasing the event-by-event fluctuations in nuclear collisions generated by the Glauber models.


\begin{thebibliography}{63}
\expandafter\ifx\csname natexlab\endcsname\relax\def\natexlab#1{#1}\fi
\expandafter\ifx\csname bibnamefont\endcsname\relax
  \def\bibnamefont#1{#1}\fi
\expandafter\ifx\csname bibfnamefont\endcsname\relax
  \def\bibfnamefont#1{#1}\fi
\expandafter\ifx\csname citenamefont\endcsname\relax
  \def\citenamefont#1{#1}\fi
\expandafter\ifx\csname url\endcsname\relax
  \def\url#1{\texttt{#1}}\fi
\expandafter\ifx\csname urlprefix\endcsname\relax\def\urlprefix{URL }\fi
\providecommand{\bibinfo}[2]{#2}
\providecommand{\eprint}[2][]{\url{#2}}

\bibitem[{\citenamefont{Alvioli
  et~al.}(2009{\natexlab{a}})\citenamefont{Alvioli, Drescher, and
  Strikman}}]{Alvioli:2009ab}
\bibinfo{author}{\bibfnamefont{M.}~\bibnamefont{Alvioli}},
  \bibinfo{author}{\bibfnamefont{H.~J.} \bibnamefont{Drescher}},
  \bibnamefont{and} \bibinfo{author}{\bibfnamefont{M.}~\bibnamefont{Strikman}},
  \bibinfo{journal}{Phys. Lett.} \textbf{\bibinfo{volume}{B680}},
  \bibinfo{pages}{225} (\bibinfo{year}{2009}{\natexlab{a}}),
  \eprint{0905.2670}.

\bibitem[{\citenamefont{Alvioli
  et~al.}(2009{\natexlab{b}})\citenamefont{Alvioli, Drescher, and
  Strikman}}]{Alvioli:page}
\bibinfo{author}{\bibfnamefont{M.}~\bibnamefont{Alvioli}},
  \bibinfo{author}{\bibfnamefont{H.~J.} \bibnamefont{Drescher}},
  \bibnamefont{and} \bibinfo{author}{\bibfnamefont{M.}~\bibnamefont{Strikman}}
  (\bibinfo{year}{2009}{\natexlab{b}}),
  \bibinfo{note}{http://www.phys.psu.edu/$\sim$malvioli/eventgenerator/}.

\bibitem[{\citenamefont{Glauber}(1959)}]{glauber}
\bibinfo{author}{\bibfnamefont{R.~J.} \bibnamefont{Glauber}},
  \bibinfo{journal}{Lectures in Theoretical Physics}
  \textbf{\bibinfo{volume}{1}} (\bibinfo{year}{1959}),
  \bibinfo{note}{{Interscience, New York}}.

\bibitem[{\citenamefont{Czyz and Maximon}(1969)}]{Czyz:1969jg}
\bibinfo{author}{\bibfnamefont{W.}~\bibnamefont{Czyz}} \bibnamefont{and}
  \bibinfo{author}{\bibfnamefont{L.~C.} \bibnamefont{Maximon}},
  \bibinfo{journal}{Annals Phys.} \textbf{\bibinfo{volume}{52}},
  \bibinfo{pages}{59} (\bibinfo{year}{1969}).

\bibitem[{\citenamefont{Baym et~al.}(1995)\citenamefont{Baym, Blattel,
  Frankfurt, Heiselberg, and Strikman}}]{Baym:1995cz}
\bibinfo{author}{\bibfnamefont{G.}~\bibnamefont{Baym}},
  \bibinfo{author}{\bibfnamefont{B.}~\bibnamefont{Blattel}},
  \bibinfo{author}{\bibfnamefont{L.~L.} \bibnamefont{Frankfurt}},
  \bibinfo{author}{\bibfnamefont{H.}~\bibnamefont{Heiselberg}},
  \bibnamefont{and} \bibinfo{author}{\bibfnamefont{M.}~\bibnamefont{Strikman}},
  \bibinfo{journal}{Phys. Rev.} \textbf{\bibinfo{volume}{C52}},
  \bibinfo{pages}{1604} (\bibinfo{year}{1995}), \eprint{nucl-th/9502038}.

\bibitem[{\citenamefont{Heiselberg}(2001)}]{Heiselberg:2000fk}
\bibinfo{author}{\bibfnamefont{H.}~\bibnamefont{Heiselberg}},
  \bibinfo{journal}{Phys. Rept.} \textbf{\bibinfo{volume}{351}},
  \bibinfo{pages}{161} (\bibinfo{year}{2001}), \eprint{nucl-th/0003046}.

\bibitem[{\citenamefont{Wang and Gyulassy}(1991)}]{Wang:1991hta}
\bibinfo{author}{\bibfnamefont{X.-N.} \bibnamefont{Wang}} \bibnamefont{and}
  \bibinfo{author}{\bibfnamefont{M.}~\bibnamefont{Gyulassy}},
  \bibinfo{journal}{Phys. Rev.} \textbf{\bibinfo{volume}{D44}},
  \bibinfo{pages}{3501} (\bibinfo{year}{1991}).

\bibitem[{\citenamefont{Werner}(1988)}]{Werner:1988jr}
\bibinfo{author}{\bibfnamefont{K.}~\bibnamefont{Werner}},
  \bibinfo{journal}{Phys. Lett.} \textbf{\bibinfo{volume}{B208}},
  \bibinfo{pages}{520} (\bibinfo{year}{1988}).

\bibitem[{\citenamefont{Broniowski
  et~al.}(2009{\natexlab{a}})\citenamefont{Broniowski, Rybczynski, and
  Bozek}}]{Broniowski:2007nz}
\bibinfo{author}{\bibfnamefont{W.}~\bibnamefont{Broniowski}},
  \bibinfo{author}{\bibfnamefont{M.}~\bibnamefont{Rybczynski}},
  \bibnamefont{and} \bibinfo{author}{\bibfnamefont{P.}~\bibnamefont{Bozek}},
  \bibinfo{journal}{Comput. Phys. Commun.} \textbf{\bibinfo{volume}{180}},
  \bibinfo{pages}{69} (\bibinfo{year}{2009}{\natexlab{a}}), \eprint{0710.5731}.

\bibitem[{\citenamefont{Alver et~al.}(2008)\citenamefont{Alver, Baker,
  Loizides, and Steinberg}}]{Alver:2008aq}
\bibinfo{author}{\bibfnamefont{B.}~\bibnamefont{Alver}},
  \bibinfo{author}{\bibfnamefont{M.}~\bibnamefont{Baker}},
  \bibinfo{author}{\bibfnamefont{C.}~\bibnamefont{Loizides}}, \bibnamefont{and}
  \bibinfo{author}{\bibfnamefont{P.}~\bibnamefont{Steinberg}}
  (\bibinfo{year}{2008}), \eprint{0805.4411}.

\bibitem[{\citenamefont{Miller et~al.}(2007)\citenamefont{Miller, Reygers,
  Sanders, and Steinberg}}]{Miller:2007ri}
\bibinfo{author}{\bibfnamefont{M.~L.} \bibnamefont{Miller}},
  \bibinfo{author}{\bibfnamefont{K.}~\bibnamefont{Reygers}},
  \bibinfo{author}{\bibfnamefont{S.~J.} \bibnamefont{Sanders}},
  \bibnamefont{and}
  \bibinfo{author}{\bibfnamefont{P.}~\bibnamefont{Steinberg}},
  \bibinfo{journal}{Ann. Rev. Nucl. Part. Sci.} \textbf{\bibinfo{volume}{57}},
  \bibinfo{pages}{205} (\bibinfo{year}{2007}), \eprint{nucl-ex/0701025}.

\bibitem[{\citenamefont{Miller and Snellings}(2003)}]{Miller:2003kd}
\bibinfo{author}{\bibfnamefont{M.}~\bibnamefont{Miller}} \bibnamefont{and}
  \bibinfo{author}{\bibfnamefont{R.}~\bibnamefont{Snellings}}
  (\bibinfo{year}{2003}), \eprint{nucl-ex/0312008}.

\bibitem[{\citenamefont{Bhalerao et~al.}(2005)\citenamefont{Bhalerao, Blaizot,
  Borghini, and Ollitrault}}]{Bhalerao:2005mm}
\bibinfo{author}{\bibfnamefont{R.~S.} \bibnamefont{Bhalerao}},
  \bibinfo{author}{\bibfnamefont{J.-P.} \bibnamefont{Blaizot}},
  \bibinfo{author}{\bibfnamefont{N.}~\bibnamefont{Borghini}}, \bibnamefont{and}
  \bibinfo{author}{\bibfnamefont{J.-Y.} \bibnamefont{Ollitrault}},
  \bibinfo{journal}{Phys. Lett.} \textbf{\bibinfo{volume}{B627}},
  \bibinfo{pages}{49} (\bibinfo{year}{2005}), \eprint{nucl-th/0508009}.

\bibitem[{\citenamefont{Andrade et~al.}(2006)\citenamefont{Andrade, Grassi,
  Hama, Kodama, and Socolowski}}]{Andrade:2006yh}
\bibinfo{author}{\bibfnamefont{R.}~\bibnamefont{Andrade}},
  \bibinfo{author}{\bibfnamefont{F.}~\bibnamefont{Grassi}},
  \bibinfo{author}{\bibfnamefont{Y.}~\bibnamefont{Hama}},
  \bibinfo{author}{\bibfnamefont{T.}~\bibnamefont{Kodama}}, \bibnamefont{and}
  \bibinfo{author}{\bibfnamefont{J.}~\bibnamefont{Socolowski},
  \bibfnamefont{O.}}, \bibinfo{journal}{Phys. Rev. Lett.}
  \textbf{\bibinfo{volume}{97}}, \bibinfo{pages}{202302}
  (\bibinfo{year}{2006}), \eprint{nucl-th/0608067}.

\bibitem[{\citenamefont{Voloshin}(2006)}]{Voloshin:2006gz}
\bibinfo{author}{\bibfnamefont{S.~A.} \bibnamefont{Voloshin}}
  (\bibinfo{year}{2006}), \eprint{nucl-th/0606022}.

\bibitem[{\citenamefont{Alver et~al.}(2006)}]{Alver:2006pn}
\bibinfo{author}{\bibfnamefont{B.}~\bibnamefont{Alver}} \bibnamefont{et~al.}
  (\bibinfo{collaboration}{PHOBOS}), \bibinfo{journal}{PoS}
  \textbf{\bibinfo{volume}{CFRNC2006}}, \bibinfo{pages}{023}
  (\bibinfo{year}{2006}), \eprint{nucl-ex/0608025}.

\bibitem[{\citenamefont{Alver et~al.}(2007{\natexlab{a}})}]{Alver:2006wh}
\bibinfo{author}{\bibfnamefont{B.}~\bibnamefont{Alver}} \bibnamefont{et~al.}
  (\bibinfo{collaboration}{PHOBOS}), \bibinfo{journal}{Phys. Rev. Lett.}
  \textbf{\bibinfo{volume}{98}}, \bibinfo{pages}{242302}
  (\bibinfo{year}{2007}{\natexlab{a}}), \eprint{nucl-ex/0610037}.

\bibitem[{\citenamefont{Sorensen}(2007)}]{Sorensen:2006nw}
\bibinfo{author}{\bibfnamefont{P.}~\bibnamefont{Sorensen}}
  (\bibinfo{collaboration}{STAR}), \bibinfo{journal}{J. Phys.}
  \textbf{\bibinfo{volume}{G34}}, \bibinfo{pages}{S897} (\bibinfo{year}{2007}),
  \eprint{nucl-ex/0612021}.

\bibitem[{\citenamefont{Broniowski et~al.}(2007)\citenamefont{Broniowski,
  Bozek, and Rybczynski}}]{Broniowski:2007ft}
\bibinfo{author}{\bibfnamefont{W.}~\bibnamefont{Broniowski}},
  \bibinfo{author}{\bibfnamefont{P.}~\bibnamefont{Bozek}}, \bibnamefont{and}
  \bibinfo{author}{\bibfnamefont{M.}~\bibnamefont{Rybczynski}},
  \bibinfo{journal}{Phys. Rev.} \textbf{\bibinfo{volume}{C76}},
  \bibinfo{pages}{054905} (\bibinfo{year}{2007}), \eprint{0706.4266}.

\bibitem[{\citenamefont{Alver et~al.}(2007{\natexlab{b}})}]{Alver:2007rm}
\bibinfo{author}{\bibfnamefont{B.}~\bibnamefont{Alver}} \bibnamefont{et~al.}
  (\bibinfo{collaboration}{PHOBOS}) (\bibinfo{year}{2007}{\natexlab{b}}),
  \eprint{nucl-ex/0701049}.

\bibitem[{\citenamefont{Aguiar et~al.}(2001)\citenamefont{Aguiar, Kodama,
  Osada, and Hama}}]{Aguiar:2000hw}
\bibinfo{author}{\bibfnamefont{C.~E.} \bibnamefont{Aguiar}},
  \bibinfo{author}{\bibfnamefont{T.}~\bibnamefont{Kodama}},
  \bibinfo{author}{\bibfnamefont{T.}~\bibnamefont{Osada}}, \bibnamefont{and}
  \bibinfo{author}{\bibfnamefont{Y.}~\bibnamefont{Hama}}, \bibinfo{journal}{J.
  Phys.} \textbf{\bibinfo{volume}{G27}}, \bibinfo{pages}{75}
  (\bibinfo{year}{2001}), \eprint{hep-ph/0006239}.

\bibitem[{\citenamefont{Hama et~al.}(2008)}]{Hama:2007dq}
\bibinfo{author}{\bibfnamefont{Y.}~\bibnamefont{Hama}} \bibnamefont{et~al.},
  \bibinfo{journal}{Phys. Atom. Nucl.} \textbf{\bibinfo{volume}{71}},
  \bibinfo{pages}{1558} (\bibinfo{year}{2008}), \eprint{0711.4544}.

\bibitem[{\citenamefont{Voloshin et~al.}(2008)\citenamefont{Voloshin,
  Poskanzer, Tang, and Wang}}]{Voloshin:2007pc}
\bibinfo{author}{\bibfnamefont{S.~A.} \bibnamefont{Voloshin}},
  \bibinfo{author}{\bibfnamefont{A.~M.} \bibnamefont{Poskanzer}},
  \bibinfo{author}{\bibfnamefont{A.}~\bibnamefont{Tang}}, \bibnamefont{and}
  \bibinfo{author}{\bibfnamefont{G.}~\bibnamefont{Wang}},
  \bibinfo{journal}{Phys. Lett.} \textbf{\bibinfo{volume}{B659}},
  \bibinfo{pages}{537} (\bibinfo{year}{2008}), \eprint{0708.0800}.

\bibitem[{\citenamefont{Manly et~al.}(2006)}]{Manly:2005zy}
\bibinfo{author}{\bibfnamefont{S.}~\bibnamefont{Manly}} \bibnamefont{et~al.}
  (\bibinfo{collaboration}{PHOBOS}), \bibinfo{journal}{Nucl. Phys.}
  \textbf{\bibinfo{volume}{A774}}, \bibinfo{pages}{523} (\bibinfo{year}{2006}),
  \eprint{nucl-ex/0510031}.

\bibitem[{\citenamefont{Alt et~al.}(2007)}]{Alt:2006jr}
\bibinfo{author}{\bibfnamefont{C.}~\bibnamefont{Alt}} \bibnamefont{et~al.}
  (\bibinfo{collaboration}{NA49}), \bibinfo{journal}{Phys. Rev.}
  \textbf{\bibinfo{volume}{C75}}, \bibinfo{pages}{064904}
  (\bibinfo{year}{2007}), \eprint{nucl-ex/0612010}.

\bibitem[{\citenamefont{Broniowski
  et~al.}(2009{\natexlab{b}})\citenamefont{Broniowski, Chojnacki, and
  Obara}}]{Broniowski:2009fm}
\bibinfo{author}{\bibfnamefont{W.}~\bibnamefont{Broniowski}},
  \bibinfo{author}{\bibfnamefont{M.}~\bibnamefont{Chojnacki}},
  \bibnamefont{and} \bibinfo{author}{\bibfnamefont{L.}~\bibnamefont{Obara}},
  \bibinfo{journal}{Phys. Rev.} \textbf{\bibinfo{volume}{C80}},
  \bibinfo{pages}{051902} (\bibinfo{year}{2009}{\natexlab{b}}),
  \eprint{0907.3216}.

\bibitem[{\citenamefont{Gazdzicki and Mrowczynski}(1992)}]{Gazdzicki:1992ri}
\bibinfo{author}{\bibfnamefont{M.}~\bibnamefont{Gazdzicki}} \bibnamefont{and}
  \bibinfo{author}{\bibfnamefont{S.}~\bibnamefont{Mrowczynski}},
  \bibinfo{journal}{Z. Phys.} \textbf{\bibinfo{volume}{C54}},
  \bibinfo{pages}{127} (\bibinfo{year}{1992}).

\bibitem[{\citenamefont{Stodolsky}(1995)}]{Stodolsky:1995ds}
\bibinfo{author}{\bibfnamefont{L.}~\bibnamefont{Stodolsky}},
  \bibinfo{journal}{Phys. Rev. Lett.} \textbf{\bibinfo{volume}{75}},
  \bibinfo{pages}{1044} (\bibinfo{year}{1995}).

\bibitem[{\citenamefont{Shuryak}(1998)}]{Shuryak:1997yj}
\bibinfo{author}{\bibfnamefont{E.~V.} \bibnamefont{Shuryak}},
  \bibinfo{journal}{Phys. Lett.} \textbf{\bibinfo{volume}{B423}},
  \bibinfo{pages}{9} (\bibinfo{year}{1998}), \eprint{hep-ph/9704456}.

\bibitem[{\citenamefont{Mrowczynski}(1998)}]{Mrowczynski:1997kz}
\bibinfo{author}{\bibfnamefont{S.}~\bibnamefont{Mrowczynski}},
  \bibinfo{journal}{Phys. Lett.} \textbf{\bibinfo{volume}{B430}},
  \bibinfo{pages}{9} (\bibinfo{year}{1998}), \eprint{nucl-th/9712030}.

\bibitem[{\citenamefont{Voloshin et~al.}(1999)\citenamefont{Voloshin, Koch, and
  Ritter}}]{Voloshin:1999yf}
\bibinfo{author}{\bibfnamefont{S.~A.} \bibnamefont{Voloshin}},
  \bibinfo{author}{\bibfnamefont{V.}~\bibnamefont{Koch}}, \bibnamefont{and}
  \bibinfo{author}{\bibfnamefont{H.~G.} \bibnamefont{Ritter}},
  \bibinfo{journal}{Phys. Rev.} \textbf{\bibinfo{volume}{C60}},
  \bibinfo{pages}{024901} (\bibinfo{year}{1999}), \eprint{nucl-th/9903060}.

\bibitem[{\citenamefont{Baym and Heiselberg}(1999)}]{Baym:1999up}
\bibinfo{author}{\bibfnamefont{G.}~\bibnamefont{Baym}} \bibnamefont{and}
  \bibinfo{author}{\bibfnamefont{H.}~\bibnamefont{Heiselberg}},
  \bibinfo{journal}{Phys. Lett.} \textbf{\bibinfo{volume}{B469}},
  \bibinfo{pages}{7} (\bibinfo{year}{1999}), \eprint{nucl-th/9905022}.

\bibitem[{\citenamefont{Appelshauser et~al.}(1999)}]{Appelshauser:1999ft}
\bibinfo{author}{\bibfnamefont{H.}~\bibnamefont{Appelshauser}}
  \bibnamefont{et~al.} (\bibinfo{collaboration}{NA49}), \bibinfo{journal}{Phys.
  Lett.} \textbf{\bibinfo{volume}{B459}}, \bibinfo{pages}{679}
  (\bibinfo{year}{1999}), \eprint{hep-ex/9904014}.

\bibitem[{\citenamefont{Voloshin}(2001)}]{Voloshin:2001ei}
\bibinfo{author}{\bibfnamefont{S.~A.} \bibnamefont{Voloshin}}
  (\bibinfo{collaboration}{STAR}) (\bibinfo{year}{2001}),
  \eprint{nucl-ex/0109006}.

\bibitem[{\citenamefont{Prindle and Trainor}(2006)}]{Prindle:2006zz}
\bibinfo{author}{\bibfnamefont{D.~J.} \bibnamefont{Prindle}} \bibnamefont{and}
  \bibinfo{author}{\bibfnamefont{T.~A.} \bibnamefont{Trainor}}
  (\bibinfo{collaboration}{STAR}), \bibinfo{journal}{PoS}
  \textbf{\bibinfo{volume}{CFRNC2006}}, \bibinfo{pages}{007}
  (\bibinfo{year}{2006}).

\bibitem[{\citenamefont{Mrowczynski}(2009)}]{Mrowczynski:2009wk}
\bibinfo{author}{\bibfnamefont{S.}~\bibnamefont{Mrowczynski}},
  \bibinfo{journal}{Acta Phys. Polon.} \textbf{\bibinfo{volume}{B40}},
  \bibinfo{pages}{1053} (\bibinfo{year}{2009}), \eprint{0902.0825}.

\bibitem[{\citenamefont{Adams et~al.}(2005{\natexlab{a}})}]{Adams:2003uw}
\bibinfo{author}{\bibfnamefont{J.}~\bibnamefont{Adams}} \bibnamefont{et~al.}
  (\bibinfo{collaboration}{STAR}), \bibinfo{journal}{Phys. Rev.}
  \textbf{\bibinfo{volume}{C71}}, \bibinfo{pages}{064906}
  (\bibinfo{year}{2005}{\natexlab{a}}), \eprint{nucl-ex/0308033}.

\bibitem[{\citenamefont{Adamova et~al.}(2003)}]{Adamova:2003pz}
\bibinfo{author}{\bibfnamefont{D.}~\bibnamefont{Adamova}} \bibnamefont{et~al.}
  (\bibinfo{collaboration}{CERES}), \bibinfo{journal}{Nucl. Phys.}
  \textbf{\bibinfo{volume}{A727}}, \bibinfo{pages}{97} (\bibinfo{year}{2003}),
  \eprint{nucl-ex/0305002}.

\bibitem[{\citenamefont{Adler et~al.}(2004)}]{Adler:2003xq}
\bibinfo{author}{\bibfnamefont{S.~S.} \bibnamefont{Adler}} \bibnamefont{et~al.}
  (\bibinfo{collaboration}{PHENIX}), \bibinfo{journal}{Phys. Rev. Lett.}
  \textbf{\bibinfo{volume}{93}}, \bibinfo{pages}{092301}
  (\bibinfo{year}{2004}), \eprint{nucl-ex/0310005}.

\bibitem[{\citenamefont{Adams et~al.}(2005{\natexlab{b}})}]{Adams:2005ka}
\bibinfo{author}{\bibfnamefont{J.}~\bibnamefont{Adams}} \bibnamefont{et~al.}
  (\bibinfo{collaboration}{STAR}), \bibinfo{journal}{Phys. Rev.}
  \textbf{\bibinfo{volume}{C72}}, \bibinfo{pages}{044902}
  (\bibinfo{year}{2005}{\natexlab{b}}), \eprint{nucl-ex/0504031}.

\bibitem[{\citenamefont{Grebieszkow et~al.}(2007)}]{Grebieszkow:2007xz}
\bibinfo{author}{\bibfnamefont{K.}~\bibnamefont{Grebieszkow}}
  \bibnamefont{et~al.}, \bibinfo{journal}{PoS}
  \textbf{\bibinfo{volume}{CPOD07}}, \bibinfo{pages}{022}
  (\bibinfo{year}{2007}), \eprint{0707.4608}.

\bibitem[{\citenamefont{Anticic et~al.}(2009)}]{na49:2008vb}
\bibinfo{author}{\bibfnamefont{T.}~\bibnamefont{Anticic}} \bibnamefont{et~al.}
  (\bibinfo{collaboration}{NA49}), \bibinfo{journal}{Phys. Rev.}
  \textbf{\bibinfo{volume}{C79}}, \bibinfo{pages}{044904}
  (\bibinfo{year}{2009}), \eprint{0810.5580}.

\bibitem[{\citenamefont{Alvioli
  et~al.}(2009{\natexlab{c}})\citenamefont{Alvioli, Atti, and
  Strikman}}]{Alvioli:2009ev}
\bibinfo{author}{\bibfnamefont{M.}~\bibnamefont{Alvioli}},
  \bibinfo{author}{\bibfnamefont{C.~C.~d.} \bibnamefont{Atti}},
  \bibnamefont{and} \bibinfo{author}{\bibfnamefont{M.}~\bibnamefont{Strikman}}
  (\bibinfo{year}{2009}{\natexlab{c}}), \eprint{0912.5025}.

\bibitem[{\citenamefont{Bialas et~al.}(1976)\citenamefont{Bialas, Bleszynski,
  and Czyz}}]{Bialas:1976ed}
\bibinfo{author}{\bibfnamefont{A.}~\bibnamefont{Bialas}},
  \bibinfo{author}{\bibfnamefont{M.}~\bibnamefont{Bleszynski}},
  \bibnamefont{and} \bibinfo{author}{\bibfnamefont{W.}~\bibnamefont{Czyz}},
  \bibinfo{journal}{Nucl. Phys.} \textbf{\bibinfo{volume}{B111}},
  \bibinfo{pages}{461} (\bibinfo{year}{1976}).

\bibitem[{\citenamefont{Kharzeev and Nardi}(2001)}]{Kharzeev:2000ph}
\bibinfo{author}{\bibfnamefont{D.}~\bibnamefont{Kharzeev}} \bibnamefont{and}
  \bibinfo{author}{\bibfnamefont{M.}~\bibnamefont{Nardi}},
  \bibinfo{journal}{Phys. Lett.} \textbf{\bibinfo{volume}{B507}},
  \bibinfo{pages}{121} (\bibinfo{year}{2001}), \eprint{nucl-th/0012025}.

\bibitem[{\citenamefont{Back et~al.}(2002)}]{Back:2001xy}
\bibinfo{author}{\bibfnamefont{B.~B.} \bibnamefont{Back}} \bibnamefont{et~al.}
  (\bibinfo{collaboration}{PHOBOS}), \bibinfo{journal}{Phys. Rev.}
  \textbf{\bibinfo{volume}{C65}}, \bibinfo{pages}{031901}
  (\bibinfo{year}{2002}), \eprint{nucl-ex/0105011}.

\bibitem[{\citenamefont{Back et~al.}(2004)}]{Back:2004dy}
\bibinfo{author}{\bibfnamefont{B.~B.} \bibnamefont{Back}} \bibnamefont{et~al.}
  (\bibinfo{collaboration}{PHOBOS}), \bibinfo{journal}{Phys. Rev.}
  \textbf{\bibinfo{volume}{C70}}, \bibinfo{pages}{021902}
  (\bibinfo{year}{2004}), \eprint{nucl-ex/0405027}.

\bibitem[{\citenamefont{Bozek}(2005)}]{Bozek:2005eu}
\bibinfo{author}{\bibfnamefont{P.}~\bibnamefont{Bozek}}, \bibinfo{journal}{Acta
  Phys. Polon.} \textbf{\bibinfo{volume}{B36}}, \bibinfo{pages}{3071}
  (\bibinfo{year}{2005}), \eprint{nucl-th/0506037}.

\bibitem[{\citenamefont{Werner}(2007)}]{Werner:2007bf}
\bibinfo{author}{\bibfnamefont{K.}~\bibnamefont{Werner}},
  \bibinfo{journal}{Phys. Rev. Lett.} \textbf{\bibinfo{volume}{98}},
  \bibinfo{pages}{152301} (\bibinfo{year}{2007}), \eprint{0704.1270}.

\bibitem[{\citenamefont{Manninen and Becattini}(2008)}]{Manninen:2008mg}
\bibinfo{author}{\bibfnamefont{J.}~\bibnamefont{Manninen}} \bibnamefont{and}
  \bibinfo{author}{\bibfnamefont{F.}~\bibnamefont{Becattini}},
  \bibinfo{journal}{Phys. Rev.} \textbf{\bibinfo{volume}{C78}},
  \bibinfo{pages}{054901} (\bibinfo{year}{2008}), \eprint{0806.4100}.

\bibitem[{\citenamefont{Bialas et~al.}(1977)\citenamefont{Bialas, Czyz, and
  Furmanski}}]{Bialas:1977en}
\bibinfo{author}{\bibfnamefont{A.}~\bibnamefont{Bialas}},
  \bibinfo{author}{\bibfnamefont{W.}~\bibnamefont{Czyz}}, \bibnamefont{and}
  \bibinfo{author}{\bibfnamefont{W.}~\bibnamefont{Furmanski}},
  \bibinfo{journal}{Acta Phys. Polon.} \textbf{\bibinfo{volume}{B8}},
  \bibinfo{pages}{585} (\bibinfo{year}{1977}).

\bibitem[{\citenamefont{Bialas et~al.}(1982)\citenamefont{Bialas, Czyz, and
  Lesniak}}]{Bialas:1980zw}
\bibinfo{author}{\bibfnamefont{A.}~\bibnamefont{Bialas}},
  \bibinfo{author}{\bibfnamefont{W.}~\bibnamefont{Czyz}}, \bibnamefont{and}
  \bibinfo{author}{\bibfnamefont{L.}~\bibnamefont{Lesniak}},
  \bibinfo{journal}{Phys. Rev.} \textbf{\bibinfo{volume}{D25}},
  \bibinfo{pages}{2328} (\bibinfo{year}{1982}).

\bibitem[{\citenamefont{Eremin and Voloshin}(2003)}]{Eremin:2003qn}
\bibinfo{author}{\bibfnamefont{S.}~\bibnamefont{Eremin}} \bibnamefont{and}
  \bibinfo{author}{\bibfnamefont{S.}~\bibnamefont{Voloshin}},
  \bibinfo{journal}{Phys. Rev.} \textbf{\bibinfo{volume}{C67}},
  \bibinfo{pages}{064905} (\bibinfo{year}{2003}), \eprint{nucl-th/0302071}.

\bibitem[{\citenamefont{Bialas and Bzdak}(2007)}]{Bialas:2006kw}
\bibinfo{author}{\bibfnamefont{A.}~\bibnamefont{Bialas}} \bibnamefont{and}
  \bibinfo{author}{\bibfnamefont{A.}~\bibnamefont{Bzdak}},
  \bibinfo{journal}{Phys. Lett.} \textbf{\bibinfo{volume}{B649}},
  \bibinfo{pages}{263} (\bibinfo{year}{2007}), \eprint{nucl-th/0611021}.

\bibitem[{\citenamefont{Bialas and Bzdak}(2008)}]{Bialas:2007eg}
\bibinfo{author}{\bibfnamefont{A.}~\bibnamefont{Bialas}} \bibnamefont{and}
  \bibinfo{author}{\bibfnamefont{A.}~\bibnamefont{Bzdak}},
  \bibinfo{journal}{Phys. Rev.} \textbf{\bibinfo{volume}{C77}},
  \bibinfo{pages}{034908} (\bibinfo{year}{2008}), \eprint{0707.3720}.

\bibitem[{\citenamefont{Bzdak}(2008)}]{Bzdak:2008gw}
\bibinfo{author}{\bibfnamefont{A.}~\bibnamefont{Bzdak}}, \bibinfo{journal}{Acta
  Phys. Polon.} \textbf{\bibinfo{volume}{B39}}, \bibinfo{pages}{1977}
  (\bibinfo{year}{2008}), \eprint{0807.1389}.

\bibitem[{\citenamefont{Tavares et~al.}(2007)\citenamefont{Tavares, Drescher,
  and Kodama}}]{Tavares:2007mu}
\bibinfo{author}{\bibfnamefont{B.~M.} \bibnamefont{Tavares}},
  \bibinfo{author}{\bibfnamefont{H.~J.} \bibnamefont{Drescher}},
  \bibnamefont{and} \bibinfo{author}{\bibfnamefont{T.}~\bibnamefont{Kodama}},
  \bibinfo{journal}{Braz. J. Phys.} \textbf{\bibinfo{volume}{37}},
  \bibinfo{pages}{41} (\bibinfo{year}{2007}), \eprint{hep-ph/0702224}.

\bibitem[{\citenamefont{Albrecht et~al.}(1989)}]{Albrecht:1989kh}
\bibinfo{author}{\bibfnamefont{R.}~\bibnamefont{Albrecht}} \bibnamefont{et~al.}
  (\bibinfo{collaboration}{WA80}), \bibinfo{journal}{Z. Phys.}
  \textbf{\bibinfo{volume}{C45}}, \bibinfo{pages}{31} (\bibinfo{year}{1989}).

\bibitem[{\citenamefont{Bachler et~al.}(1993)}]{Bachler:1992jp}
\bibinfo{author}{\bibfnamefont{J.}~\bibnamefont{Bachler}} \bibnamefont{et~al.}
  (\bibinfo{collaboration}{NA35}), \bibinfo{journal}{Z. Phys.}
  \textbf{\bibinfo{volume}{C57}}, \bibinfo{pages}{541} (\bibinfo{year}{1993}).

\bibitem[{\citenamefont{Aggarwal et~al.}(2002)}]{Aggarwal:2001aa}
\bibinfo{author}{\bibfnamefont{M.~M.} \bibnamefont{Aggarwal}}
  \bibnamefont{et~al.} (\bibinfo{collaboration}{WA98}), \bibinfo{journal}{Phys.
  Rev.} \textbf{\bibinfo{volume}{C65}}, \bibinfo{pages}{054912}
  (\bibinfo{year}{2002}), \eprint{nucl-ex/0108029}.

\bibitem[{\citenamefont{Adare et~al.}(2008)}]{Adare:2008ns}
\bibinfo{author}{\bibfnamefont{A.}~\bibnamefont{Adare}} \bibnamefont{et~al.}
  (\bibinfo{collaboration}{PHENIX}), \bibinfo{journal}{Phys. Rev.}
  \textbf{\bibinfo{volume}{C78}}, \bibinfo{pages}{044902}
  (\bibinfo{year}{2008}), \eprint{0805.1521}.

\bibitem[{\citenamefont{Rybczynski et~al.}(2008)\citenamefont{Rybczynski,
  Broniowski, and Bozek}}]{Rybczynski:2008zg}
\bibinfo{author}{\bibfnamefont{M.}~\bibnamefont{Rybczynski}},
  \bibinfo{author}{\bibfnamefont{W.}~\bibnamefont{Broniowski}},
  \bibnamefont{and} \bibinfo{author}{\bibfnamefont{P.}~\bibnamefont{Bozek}},
  \bibinfo{journal}{Acta Phys. Polon.} \textbf{\bibinfo{volume}{B39}},
  \bibinfo{pages}{1725} (\bibinfo{year}{2008}), \eprint{0803.4294}.

\bibitem[{\citenamefont{Rybczynski and Wlodarczyk}(2005)}]{Rybczynski:2004zi}
\bibinfo{author}{\bibfnamefont{M.}~\bibnamefont{Rybczynski}} \bibnamefont{and}
  \bibinfo{author}{\bibfnamefont{Z.}~\bibnamefont{Wlodarczyk}},
  \bibinfo{journal}{J. Phys. Conf. Ser.} \textbf{\bibinfo{volume}{5}},
  \bibinfo{pages}{238} (\bibinfo{year}{2005}), \eprint{nucl-th/0408023}.

\end{thebibliography}

\end{document}